\documentclass{article}
\usepackage{spconf,amsmath,graphicx,url,subfig, cite,booktabs}
\usepackage{graphicx}
\usepackage{amsfonts,amssymb}
\usepackage{xcolor}
\usepackage{lipsum}

\title{RaD-Net: A Repairing and Denoising Network for Speech Signal Improvement}
\name{
\begin{tabular}{c}
\it Mingshuai Liu$^{1,2}$, Zhuangqi Chen$^2$, Xiaopeng Yan$^1$, Yuanjun Lv$^1$, Xianjun Xia$^2$, \\ Chuanzeng Huang$^2$, 
Yijian Xiao$^2$, Lei Xie$^{1*}$\thanks{* Corresponding author.}
\end{tabular}
}
\address{
  $^1$Audio, Speech and Language Processing Group (ASLP@NPU), School of Software, \\ Northwestern Polytechnical University, Xi'an, China\\
  $^2$ByteDance, China
  \vspace{-0.15em}
  }
\begin{document}
%
\maketitle
\begin{abstract}
\vspace{-0.35em}
This paper introduces our repairing and denoising network (RaD-Net) for the ICASSP 2024 Speech Signal Improvement (SSI) Challenge. We extend our previous framework based on a two-stage network and propose an upgraded model. Specifically, we replace the repairing network with COM-Net from TEA-PSE. In addition, multi-resolution discriminators and multi-band discriminators are adopted in the training stage. Finally, we use a three-step training strategy to optimize our model. We submit two models with different sets of parameters to meet the RTF requirement of the two tracks. According to the official results, the proposed systems rank 2nd in track 1 and 3rd in track 2\footnote{Demo page is available at https://github.com/mishliu/RaD-Net}.
\end{abstract}
\vspace{-4pt}
\vspace{-1.0em}
\section{Introduction}
\label{sec:intro}
\vspace{-0.9em}

Speech communication systems play a crucial role in daily life and work. However, during audio communication, speech signals may suffer from multiple distortions including coloration, discontinuity, loudness, noisiness, and reverberation, which damage speech signal quality. The ICASSP 2024 SSI Challenge\footnote{https://www.microsoft.com/en-us/research/academic-program/speech-signal-improvement-challenge-icassp-2024} focuses on improving send speech signal quality in mainstream communication systems.

In the ICASSP 2023 SSI Challenge, we propose a two-stage neural network~\cite{sig2023_new} to address the aforementioned problem. Specifically, we first employ the repairing network for frequency response distortions, isolated and nonstationary distortions, and loudness issues. Then the denoising network is applied to remove noise, reverberation, and artifacts. This year, we advance our previous two-stage neural network and propose an upgraded system. First, we replace the repairing network with COM-Net from TEA-PSE~\cite{tea2_new}. Meanwhile, the repairing network is also applied for preliminary denoising and dereverberation in the proposed 
\begin{figure}[htbp!]
	\begin{minipage}{1.0\linewidth}
		\centering
		\includegraphics[width=1.0\linewidth]{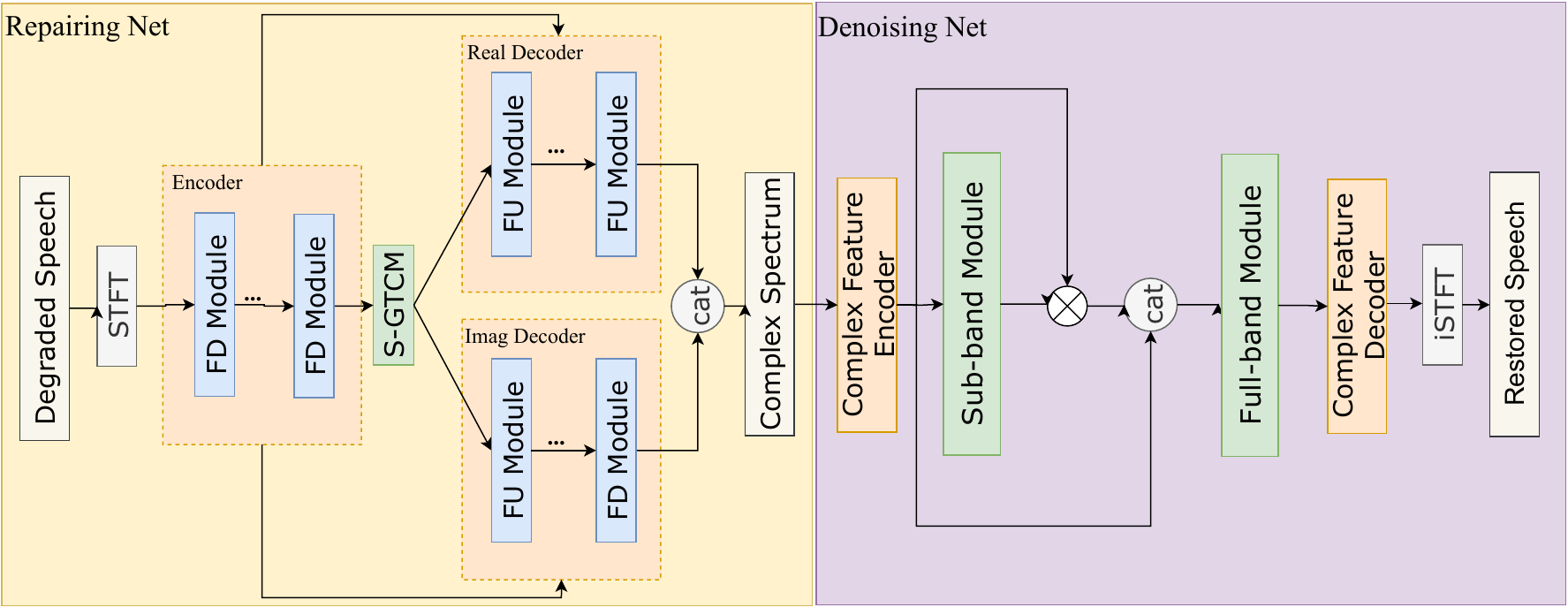}
		\label{f2_2}
	\end{minipage}
    \vspace{-2.2em}
    \caption{Architecture of RaD-Net.}
    \vspace{-2.1em}
    \label{f2}
\end{figure}
systems. Second, to improve speech naturalness, we introduce multi-resolution discriminators~\cite{mfd_new} and multi-band discriminators~\cite{mfd_new} in the training stage. Finally, we use a more effective three-step training strategy to accelerate the training convergence.


\vspace{-1.4em}
\section{Method}
\label{sec:format}

\vspace{-0.9em}
\subsection{Repairing Net}
\label{ssec:subhead}
\vspace{-0.8em}
TEA-PSE~\cite{tea2_new}, a top-ranking TSE model in recent DNS challenges, shows impressive performance in personalized speech enhancement. TEA-PSE includes MAG-Net and COM-Net to handle magnitude and complex-valued features respectively. To recover the phase information from the distorted speech, we replace our previous GateDCCRN with COM-Net.

As shown in Fig.1, the repairing network adopts an encoder-decoder architecture. The encoder consists of three frequency down-sampling (FD) layers, while the decoders are stacked by three frequency up-sampling (FU) layers. The FD layer starts with a gated convolution layer (GateConv), followed by a cumulative layer norm (cLN), PReLU, and time-frequency convolution module (TFCM). The FU layer uses a mirror structure and replaces the GateConv with a transposed gated convolutional layer (TrGateConv). 
Between the encoder and decoder, we apply four stacked gated temporal convolutional modules (S-GTCM) for temporal modeling.

\vspace{-1.6em}
\subsection{Denoising Net}
\label{ssec:subhead}
\vspace{-0.8em}
In the ICASSP 2023 SSI Challenge, we employed a variant of S-DCCRN~\cite{sdccrn} named S-DCCSN as the denoising network. As shown in Fig.1, a sub-band module and a full-band module are cascaded to process local and global frequency information, respectively. This year, to balance the computational cost and system performance, we simplify the S-DCCSN by decreasing the number of channels and replacing convolution layers with depthwise separable convolution layers.

\vspace{-1.4em}
\subsection{Training Strategy}
\label{ssec:subhead}
\vspace{-0.8em}
Inspired by the superior performance of adversarial learning, we introduce multi-resolution discriminators and multi-band discriminators in the training stage. Each discriminator is comprised of multiple 2D convolution layers. 
The input of multi-resolution discriminators is magnitude, while the input of multi-band discriminators is complex spectrums.

To accelerate the training convergence, we use a more effective three-step training strategy. We first train the repairing network with discriminators and freeze this pre-trained model. Next, we pre-train the denoising network on denoising and dereverberation tasks. Note that discriminators are not used in this step. Finally, we load the pre-trained repairing and denoising network and finetune only the parameters of the denoising network with discriminators. 

\vspace{-1.4em}
\subsection{Loss Function}
\label{ssec:subhead}
\vspace{-0.8em}
We train the repairing network with $\mathcal{L}_{\text{1}}$, which is composed of  the spectral convergence loss $\mathcal{L}_{\text{sc}}$~\cite{scloss}, the L1 loss of the logarithmic magnitude, the asymmetric loss $\mathcal{L}_{\text{log-mag}}$, the generator loss $\mathcal{L}_{\text{G}}$, and the feature matching loss $\mathcal{L}_{\text{FM}}$. For the denoising network, we first use $\mathcal{L}_{\text{2-pre}}$ which consists of the scale-invariant signal-to-noise ratio (SI-SNR) loss $\mathcal{L}_{\text{si-snr}}$ and power-law compressed loss $\mathcal{L}_{\text{plc}}$ in the pre-training stage. Then we further incorporate $\mathcal{L}_{\text{G}}$ and $\mathcal{L}_{\text{FM}}$ into $\mathcal{L}_{\text{2-pre}}$ and obtain $\mathcal{L}_{\text{2}}$ to finetune the denoising network.
\vspace{-0.8em}
\begin{equation}
\begin{aligned}
    & \mathcal{L}_{\text{1}}=\mathcal{L}_{\text{sc}}+\mathcal{L}_{\text{log-mag}}+\mathcal{L}_{\text{G}}+\mathcal{L}_{\text{FM}}, \\
    & \mathcal{L}_{\text{2-pre}}=\mathcal{L}_{\text{si-snr}}+\mathcal{L}_{\text{plc}}, \\
    & \mathcal{L}_{\text{2}}=\mathcal{L}_{\text{si-snr}}+\mathcal{L}_{\text{plc}}+\mathcal{L}_{\text{G}}+\mathcal{L}_{\text{FM}}.
\end{aligned}
\end{equation}

\vspace{-2.0em}
\section{Experiment}
\label{sec:typestyle}
\vspace{-1.0em}

\subsection{Dataset}
\label{ssec:subhead}
\vspace{-0.8em}
We use subsets from the DNS5 dataset~\cite{dns5_new} as the clean speech set and noise set. In this challenge, a demo data synthesizer is released to generate distorted audio from clean audio. Besides this synthesizer, we also use OPUS, AAC, AMR, and a private noise suppressor (NS) for data simulation. In total, a 1200-hour dataset is finally used for training and evaluation.

\vspace{-1.3em}
\subsection{Training Setup}
\label{ssec:subhead}
\vspace{-0.8em}
We submitted two versions of RaD-Net in track 1 and track 2, respectively. The distinction lies in the number of channels for S-DCCSN. The window length and frameshift are 20ms and 10ms, respectively. The STFT length is 960. The number of channels, stride, and kernel size of GConv and TrGConv are 64, (4, 1), and (5, 1), respectively. One TFCM contains 3 depthwise dilated convolution layers with a dilation rate of \{1,2,4\} and a kernel size of (3, 5). One S-GTCM contains 4 GTCM layers with a kernel size of 5 for dilated convolution layers and a dilation rate of \{1,2,5,9\}, respectively. In track 1, the number of channels for the sub-band module and the full-band module in S-DCCSN is \{32, 32, 32, 32, 64, 64\}, while in track 2, it is \{64, 64, 64, 64, 64, 128\}. Models are optimized by AdamW with an initial learning rate of 0.0002.

\subsection{Results and Analysis}
\vspace{-0.7em}
To prove the effectiveness of our model, we conduct ablation experiments using the 1200-hour dataset. 
As shown in Table 1, we can observe that the repairing network achieves a significant improvement in SIG, BAK, and OVRL by replacing GateDCCRN with COM-Net. In addition, DNSMOS results in Table 1 show that the adversarial training strategy using multi-resolution discriminators and multi-band discriminators can effectively improve the signal improvement ability of both the repairing network and the denoising network. Table 2 shows the subjective results of the RaD-Net on the official test set. We can see that speech quality is clearly improved for all types of distortions. In track 1, the RTF is 0.45 tested on Intel(R) Xeon(R) CPU E5-2678 v3 2.4GHz using a single thread, while it is 0.60 in track 2.

\vspace{-0.7em}
\label{ssec:subhead}
\begin{table}[htbp]
\centering
\small
 \caption{DNSMOS for ablation models on the blind set. ``S1", ``S2", and ``D" denote the repairing network in stage 1, the denoising network in stage 2, and the discriminators.}
 \vspace{-8pt}
\setlength{\tabcolsep}{0.6mm}
 \label{tab:dnsmos}
  \resizebox{\linewidth}{!}{
\begin{tabular}{@{}lcccccccccccccccccccccc@{}}
\toprule
    Model & & & & & & Para.~(M) & & & & & SIG & & & & & & BAK & & & & & OVRL \\ \midrule
    Noisy & & & & & & - & & & & & 3.00 & & & & & & 3.57 & & & & & 2.62       \\ 
S1 & & & & & & 2.21 & & & & & 3.19 & & & & & & 3.87 & & & & & 2.85      \\ 
GateDCCRN with D & & & & & & 6.70 & & & & & 3.27 & & & & & & 3.99 & & & & & 2.97      \\
S1 with D & & & & & & 2.21 & & & & & 3.30 & & & & & & 4.01 & & & & & 3.02    \\
S1 with D+S2 & & & & & & 4.00 & & & & & 3.29 & & & & & & 4.09 & & & & & 3.06      \\
S1 with D+S2 with D & & & & & & 4.00 & & & & & 3.34 & & & & & & 4.10 & & & & & 3.10      \\
\bottomrule
\end{tabular}
}
\vspace{-2.2em}
\end{table}

\begin{table}[htbp]
\centering
\tiny
 \caption{Multi-dimensional subjective results on the blind set.}
 \vspace{-8pt}
\setlength{\tabcolsep}{0.6mm}
 \label{tab:corrcoef}
  \resizebox{\linewidth}{!}{
\begin{tabular}{@{}lcccccccc@{}}
\toprule
    & Track & COL & DISC & LOUD & NOISE & REVERB & SIG & OVRL \\ \midrule
Noisy & - & 3.341 & 3.695 & 3.778 & 3.214 & 3.399 & 3.049 & 2.580 \\ 
RaD-Net & 1 & 3.763 & 3.819 & 4.216 & 4.070 & 3.930 & 3.528 & 3.167 \\
RaD-Net & 2 & 3.774 & 3.817 & 4.228 & 4.051 & 3.887 & 3.517 & 3.163 \\
\bottomrule
\end{tabular}
}
\vspace{-2.4em}
\end{table}

\vspace{-0.7em}
\section{Conclusions}
\label{sec:majhead}
\vspace{-0.9em}

This paper introduces our submission to the ICASSP 2024 SSI Challenge. We advance our previous two-stage neural network by replacing the repairing network with COM-Net, introducing frequency domain discriminators in the training stage, and using a three-step training strategy. Our systems rank 2nd in track 1 and 3rd in track 2.

\footnotesize
\vspace{-0.4em}
\bibliographystyle{IEEEbib}
\let\oldbibliography\thebibliography 
\renewcommand{\thebibliography}[1]{ 
  \oldbibliography{#1}
  \setlength{\itemsep}{-1pt} 
}
\bibliography{strings,refs}

\end{document}